\documentclass[aps,prx,reprint]{revtex4-1}
\usepackage{blindtext}
\usepackage{graphicx}% Include figure files
\usepackage{xcolor}
\usepackage{dcolumn}% Align table columns on decimal point
\usepackage{bm}% bold math
\usepackage[utf8]{inputenc}

\begin{document}
	
	\preprint{APS/123-QED}
	
	\title{Spectroscopic evidence of high temperature superconductivity in VSe$_{2}$}% Force line breaks with \\}
	
	\author{Turgut Yilmaz}
	\affiliation{National Synchrotron Light Source II, Brookhaven National Lab, Upton, New York, 11973, USA}
	
	\author{Elio Vescovo}
	\affiliation{National Synchrotron Light Source II, Brookhaven National Lab, Upton, New York, 11973, USA}
	
	\author{Jerzy T. Sadowski}
	\affiliation{Center for Functional Nanomaterials, Brookhaven National Lab, Upton, New York, 11973, USA}
	
	\author{Boris Sinkovic}
	\affiliation{Department of Physics, University of Connecticut, Storrs, Connecticut, 06269, USA}

	\date{\today}% It is always \today, today,
	
	\begin{abstract}
		High-resolution angle-resolved photoemission experiments reveal subtle modifications of the surface electronic structure of VSe$_{2}$. Most remarkably, we show that superconductivity can be induced in VSe$_{2}$ by the right selection of substrate and growth parameters. Evidence for the superconducting state comes from the simultaneous detection of spectral kink, quasiparticle peak, Fermi gap, and their evolution with the temperature. The observation of Bogoliubov-like back-bending bands at low temperatures, signaling electron-hole pairing, further supports the presence of superconductivity in this system. The photoemission experiment also provides evidence for a formation of a pseudogap state at high temperatures, characterized by the progressive quenching of the quasiparticle peak feature coexisting with a persistent gap at the Fermi level, a behavior reminiscent of high-T$_{c}$ superconductors. We attributed the origin of superconductivity in these VSe$_{2}$ films to a modified Fermi surface combined with the formation of van Hove singularity points with a binding energy corresponding to the chemical potential of the system. Although T$_{c}$ cannot be accurately determined from photoemission data, observations based on the survival temperature of the quasiparticle peak suggest that T$_{c}$ could be as high as 100 $\pm$ 5 K and substantially higher than previous reports for any transition metal dichalcogenides.

	\end{abstract}

	\maketitle
	
	%\tableofcontents
	
	\section{Introduction}
	
	More than 100 years after its discovery, superconductivity remains one of the most intrigin phenomena in material science; an apparently inexhaustible source of scientific excitement and technological interest. The observation of unconventional superconductivity in (La,Ba)$_{2}$CuO$_{4}$ a few decades ago \cite{bednorz1986possible}, with their unexpectedly high-T$_{c}$, redoubled interest in this phenomenon. Once broken the taboo of ultra-low temperatures, the challenge of understanding the physics of the new superconductors became inseparably connected with the hope of achieving T$_{c}$ above ambient temperature. Nowadays, the adoption of superconducting materials as basic components in quantum computing (Qbits) has become another driving force for research in this area \cite{devoret2004superconducting}. There is therefore a constant search for new superconductors, each addition to the list contributing to a more complete picture of the phenomenon while at the same time expanding the range of potential applications. Interestingly, the two latest entries in the list of high-T$_{c}$ FeSe monolayer \cite{he2013phase} and twisted graphene bilayers \cite{cao2018unconventional} have put a new spin in the search for room temperature and /or topological superconductivity, suggesting that equally important to the synthesis of new materials, is the accurate control of growth parameters and lattice orientations in the preparation of old ones.

	Among materials’ families, layered transition metal dichalcogenides (TMDCs) are under intense scrutiny in the search for superconductivity. Although relatively simple materials, they host an impressive variety of physics, including Mott insulating state, charge density waves, and ferromagnetism \cite{manzeli20172d, choi2017recent}. This richness offers a possibility to study the interplay between superconductivity and other states of matter. In this connection, particularly remarkable is the case of  TMDC displaying topological order, another exotic state of matter which can induce Majorana fermions if it coexists with superconductivity \cite{shi2015superconductivity, li2021observation, fu2008superconducting}. This can provide a realistic material candidate for fault-tolerant quantum computing \cite{lian2018topological}. However, so far, the highest observed superconducting T$_{c}$ in TMDCs is 8.8 K reported for 2M-WS$ _{2} $ \cite{fang2019discovery}. Such a low-T$_{c}$ hinders the technological adoption of these materials.
	
	Here, we report on a comprehensive high-resolution angle-resolved photoemission (ARPES) study of  one TMDC , VSe$_{2}$. Optimally grown films display all the hallmarks of high-T$_{c}$ superconductivity: the formation of the spectral kink, the quasiparticle peak (QP), and the gap at the Fermi level. As a function of temperature, the QP is observed up to ~ 110 K while the Fermi gap persists with decreasing size up to ~150 K, indicating the presence of a pseudogap in analogy to high-T$_{c}$ superconductors. The superconducting state is further verified by the observation of Bogoliubov back-bending bands that are the standard signature of electron-hole pairing in the superconducting state \cite{matsui2003bcs}. These spectral features are very sensitive to the films' growth condition; in particular, the substrate type / quality and the growth temperature are crucial for inducing superconductivity in VSe$_{2}$. Even though the photoemission data do not provide a direct way of determining T$_{c}$, the temperature evolution of the spectral line shape suggests that T$_{c}$ could be as high as 100 $\pm$ 5 K. This would be much higher than previously reported for any TMDCs, providing a potential material candidate with easily accessible experimental parameters to investigate the superconductivity  and its connection to the other quantum states.
	
	\begin{figure*}[t]
		\centering
		\includegraphics[width=16cm,height=11.574cm]{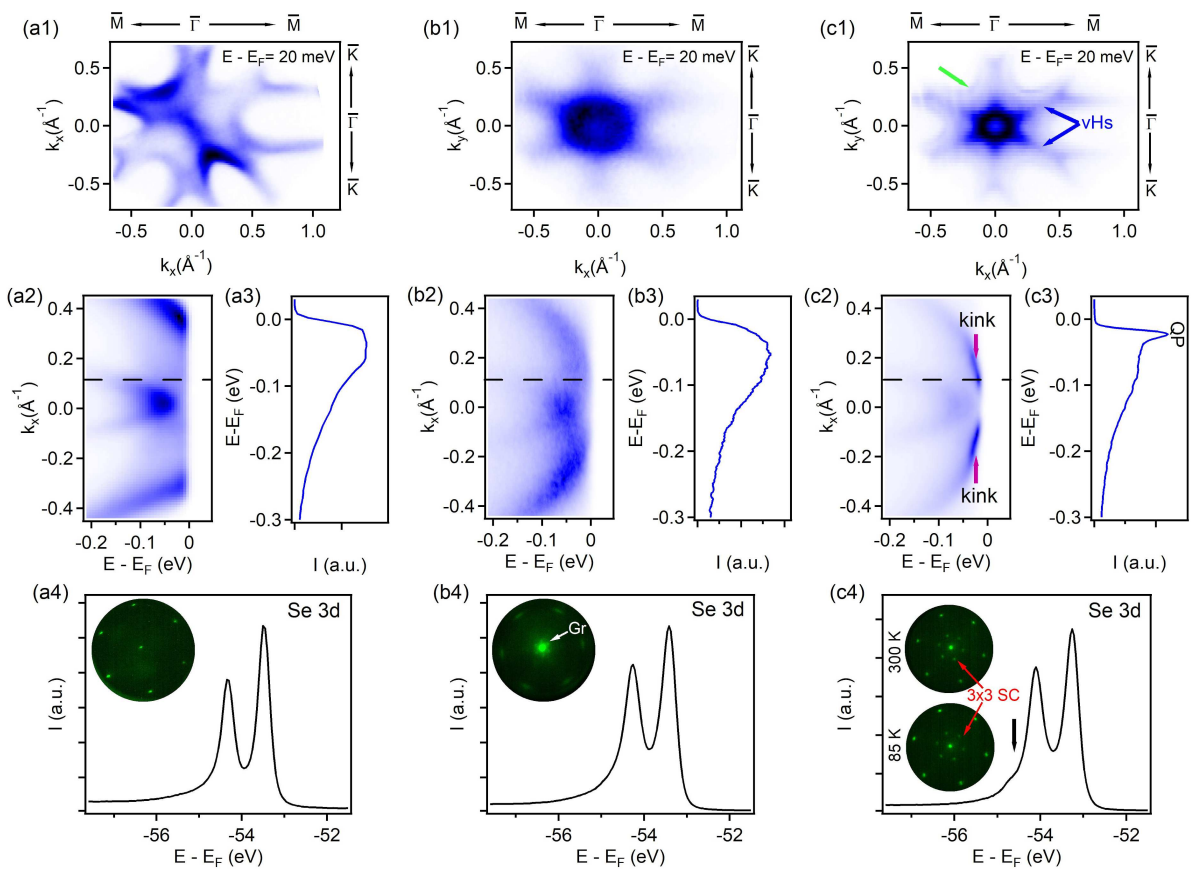}
		
		\caption{(a1) Constant energy cut at 20 meV with respect to the Fermi level obtained from a vacuum cleaved VSe$_{2}$ single crystal. (a2) Binding energy versus in-plane momentum map of the same sample taken in the $\overline{\Gamma}$ - $\overline{M}$ direction of the Brillouin zone. (a3) EDC obtained along the dashed black line in (a2). (a4) Corresponding Se 3d peak recorded with 110 eV photons at 10 K. Inset in (a4) is the $\mu$-low energy electron diffraction ($\mu$LEED) pattern of the same sample. (b1) - (b4) and (c1) - (c4) are the same as (a1) - (a4) but for 5 ML VSe$_{2}$ thin films grown on Gr/SiC and HOPG substrates, respectively. The same growth parameters are adopted for the thin films and the substrate temperatures were kept at 570 K during the growth. Unlike other two samples, the VSe2 sample grown on HOPG displays a 3 x 3 superstructure $\mu$LEED pattern at 300 K and 85 K as indicated in the inset of (c4). The blue arrow in (c1) shows the van Hove singularity points while the green arrow points to a Fermi arc. Purple arrow in (c3) marks the spectral kink. The black arrow in (c4) marks a new spectral feature as a shoulder in the higher binding energy side of  Se 3d for the film grown on HOPG. This feature is absent in other two samples. All the ARPES maps are collected with 50 eV photons at 10 K.}
		
	\end{figure*}
	
	\begin{figure*}[t]
		\centering
		\includegraphics[width=16cm,height=8.251cm]{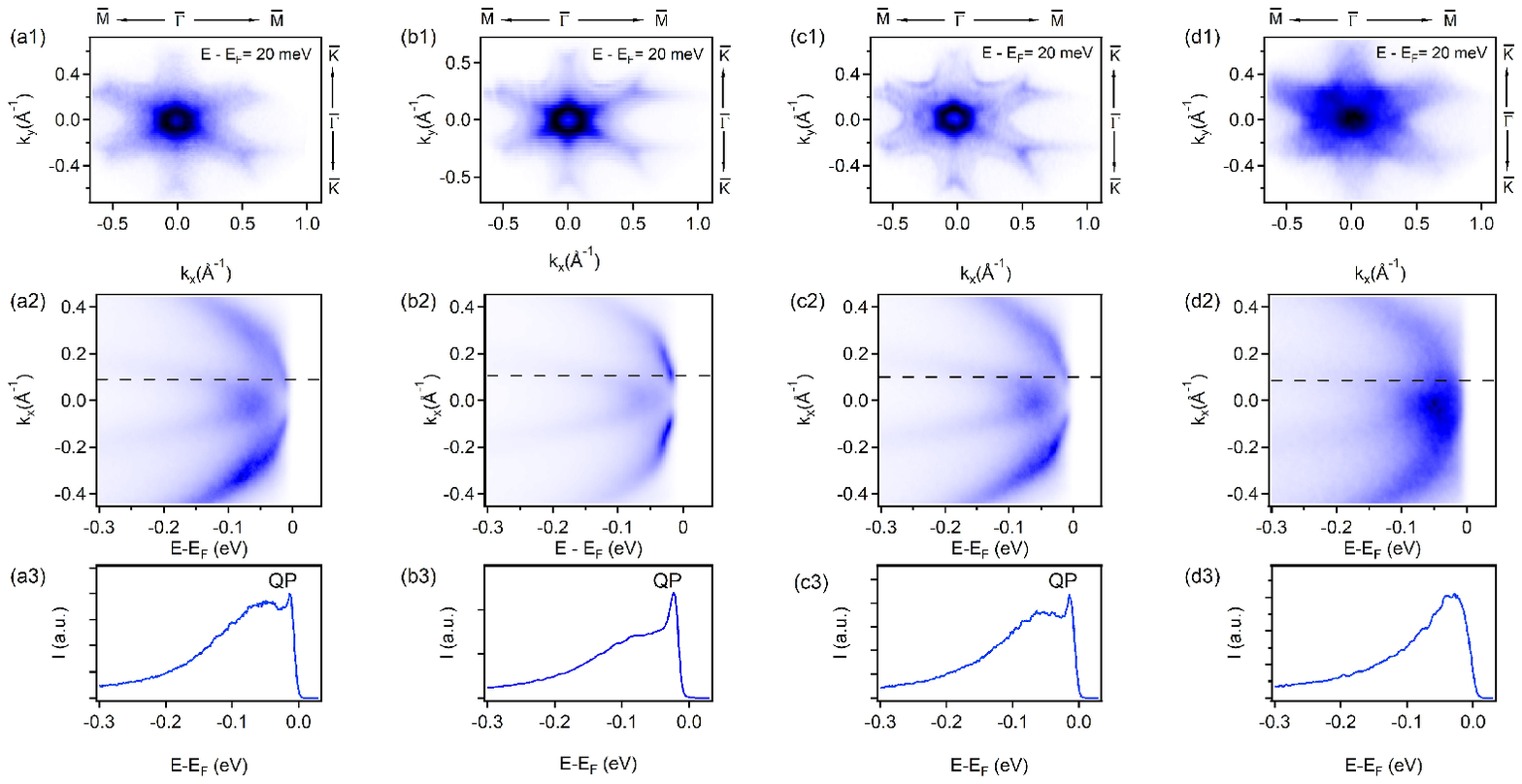}
		\caption{Top panels: constant energy cuts, 20 meV below the Fermi level. Middle panels: ARPES band dispersions along the$\overline{\Gamma}$ - $\overline{M}$ high symmetry direction of the Brillouin zone. Lower panels: EDCs along the dashed line in dispersion spectra. All films were grown in identical conditions except for the substrate temperatures set to 550 K, 570 K, 600 K, and 650 K from left to right. Spectra were measured at 10 K with 50 eV photons from the samples grown on HOPG substrates.}
	\end{figure*}

	\section{Impact of the growth method and substrate on the electronic structure}

	To investigate the impact of the growth method on the electronic and structural properties, we compare various form of VSe$_{2}$: bulk crystal vs. epitaxial films deposited on different substrates. These comparisons are summarized in Fig. 1, presenting the ARPES maps,  $\mu$LEED patterns, and Se 3d core-level spectra. The left column is for the bulk single crystals synthesized by the flux zone method. The central and right columns are for two epitaxial films prepared by molecular beam epitaxy (MBE) under nominally identical conditions on Gr/SiC and HOPG, respectively. At first glance, it is apparent that qualitative differences distinguish the electronic structure of the bulk sample from the thin films and, even more surprisingly, also of the two films grown on the two similar substrates. The Fermi surface of the bulk VSe$_{2}$ exhibits two distinct features: an intense emission at the $\overline{\Gamma}$-point and ellipsoidal electron pockets centered at the $\overline{M}$-points (Fig. 1(a1)). The ellipsoidal features are attributed to the V 3d band, while the emission at the zone center derives from Se 4p atomic orbitals \cite{duvjir2018emergence}. The energy dispersion of the valence bands along the $\overline{\Gamma}$ - $\overline{M}$ direction is displayed in Fig. 1(a2). The V 3d states dominate the region close to the Fermi level while the top of the Se 4p reaches the binding energy of about 0.2 eV at the $\overline{\Gamma}$-point.
	
	\begin{figure*}[t]
		\centering
		\includegraphics[width=16cm,height=9.644cm]{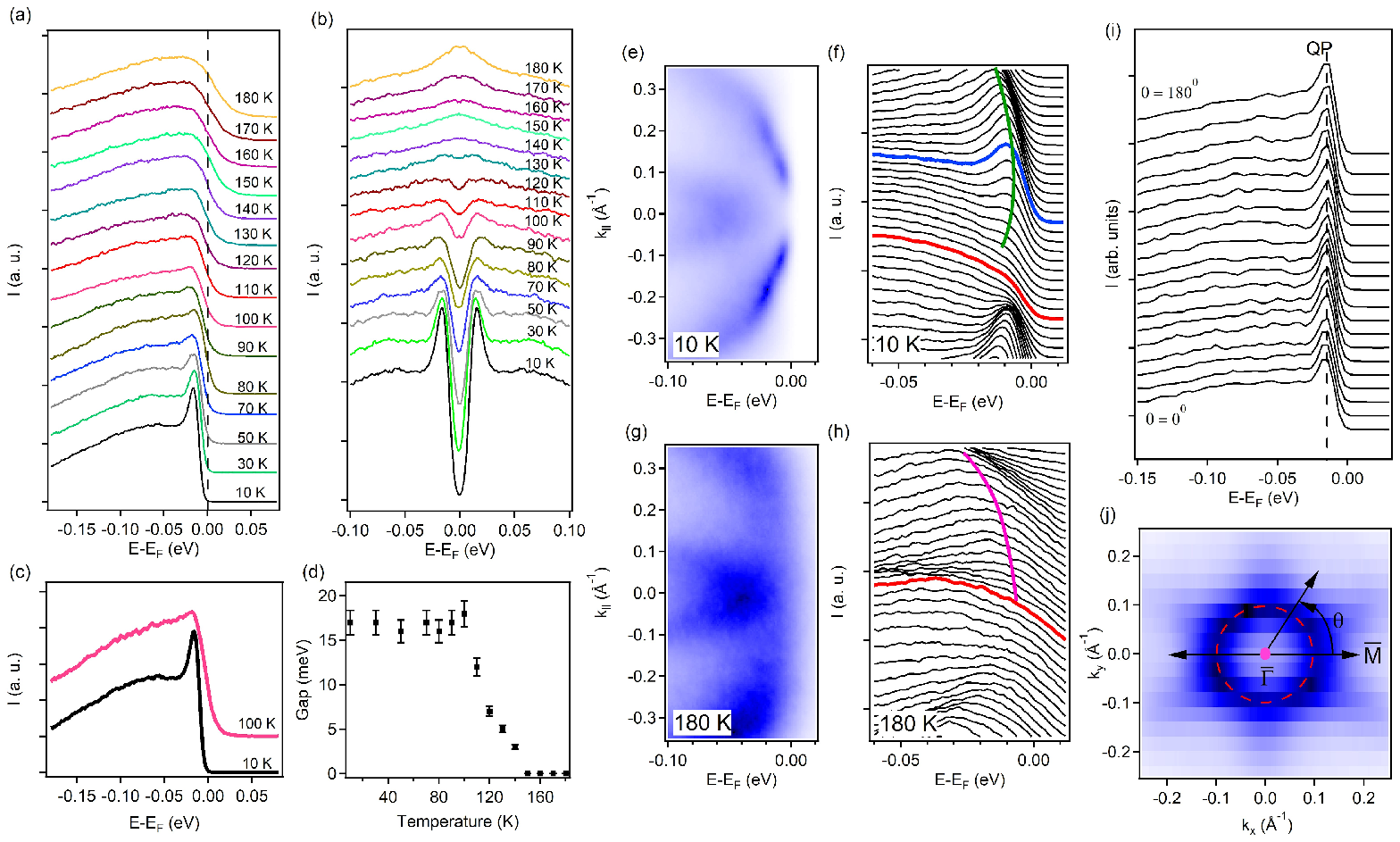}
		\caption{(a) QP-EDCs vs. temperature. (b) Symmetrized EDCs. (c) Comparison of EDCs taken at 10 K and 100 K. (d) Fermi gap vs. temperature. The gap is estimated from the symmetrized EDC given in (b). (e) and (f) Dispersion map taken at 10 K and corresponding momentum integrated EDCs. The red line and blue line in (f) mark the $\overline{\Gamma}$-point and the Fermi momentum, respectively. The green line in f represents the inward band bending. (g) and (h) are the same as (e) and (f) but taken at 180 K sample temperature. The pink line in (h) marks the nearly flat dispersion in the vicinity of the Fermi level and around the $\overline{\Gamma}$-point. (i) QP-EDCs as a function of the angle-$\theta$ in (j) where constant energy cut 20 meV below the Fermi level measured at 10 K is given. All the spectra are obtained with 50 eV photons from the sample grown on HOPG substrates at 570 K substrate temperature.}
	\end{figure*}
	
	Comparing the Fermi surface of bulk sample to that of the 5 ML VSe$_{2}$ film grown on Gr/SiC, the main difference is the formation of a hexagonal-shaped hole pocket centered at the $\overline{\Gamma}$-point. Away from the zone center, the ellipse-shaped electron pockets centered at the $\overline{M}$-point and the nearly -triangular hole-like pocket around the $\overline{K}$-point are similar although less intense than in  the bulk sample (Fig. 2(b1)). The binding energy – in-plane momentum (k$ _{\Vert} $) map of the VSe$_{2}$ film on Gr/SiC exhibits an M-shape dispersive band in the vicinity of the Fermi level (Fig. 1(b2)). Energy distribution curves (EDCs) of both samples presented in Fig. 1(a3) and 1(b3), respectively, display a similar density of states with a broad bump located in the vicinity of the Fermi level without any spectral anomaly. The electronic structures presented here for the single crystal and the thin film grown on Gr/SiC are consistent with the literature \cite{duvjir2018emergence, feng2018electronic, coelho2019charge}.

	The case of a 5 ML VSe$_{2}$ film grown on a HOPG substrate is the most interesting one. New structures develop compared to both the bulk samples and the films on graphene. The center of the Fermi surface assumes a star-like shape and intense triangular pockets surround the $\overline{K}$points (Fig. 1(c1)). The vertices of these two features connect along the $\overline{\Gamma}$ - $\overline{M}$ directions at about 1/3 of the way, to form van Hove singularity (vHs) points. The six-fold symmetric ellipsoidal electron pockets centered at the $\overline{M}$-points are also well resolved in this sample but and are now delimited by extremely sharp momentum arcs (green arrow in Fig. 1(c1)). Finally, a well-defined circular electron pocket is now present close to the $\overline{\Gamma}$ point (Fig. 1(c1)). These states are concentrated in a narrow energy window of about 25 meV in the vicinity of the Fermi level, possibly resulting from fine tuning of the binding energy of vHs points. Relevant information on the unique electronic structure of this sample is contained in the dispersion map (Fig. 1(c2)). First, the V 3d derived bands approaching the Fermi level exhibit a spectral kink (see red arrows in Fig. 1(c2)). Second, an intense QP peak is now present in the EDC, followed by a dip and a hump of decreasing intensity towards higher binding energy (Fig. 1(c3)). Such spectral line shape is characteristic of the superconducting state \cite{dessau1991anomalous}. Therefore, our observations suggests that VSe$_{2}$ can be turned into a superconductor when grown on a HOPG substrate under proper conditions.

	To further investigate the origin of the spectral differences in these samples, we report $\mu$LEED patterns taken at 300 K and the Se 3d peaks for the corresponding samples. The diffraction pattern of single-crystal VSe$_{2}$ displays sharp spots, forming the expected three-fold diffraction pattern (inset in Fig. 1(a4)). For the sample grown on the Gr substrate, the diffraction spots are broader and elongated along the tangential direction, indicating the presence of the multiple domains (inset in Fig. 1(b4)). This broadening is consistent with what is observed in ARPES. Furthermore, Se 3d peaks of both samples show a single doublet line shape, again somewhat broader in the case of the film on Gr (Fig. 1(a4) and 1(b4)). Corresponding to the unique features observed in the ARPES data, the $\mu$LEED pattern for the VSe$_{2}$ sample grown on HOPG exhibits a sharp 3 x 3 superstructure at 300 K and 85 K, and the Se 3d doublet has an additional component at higher binding energy, marked with a black arrow in Fig. 1(c4). These features are possibly due to V impurity intercalation as seen in a variety of similar compounds \cite{bonilla2020compositional}. We should also emphasize that impurity intercalation can affect the chemical potential of the system and consequently modify the binding energy of the vHs points. This mechanism could explain the suppression of the charge density waves in these films and enhance the superconducting $T_{c}$ \cite{chikina2020turning}. Furthermore, absence of the any difference between the low temperature and room temperature $\mu$LEED shows that the emergent QP peak, spectral kink, and vHs points in VSe$_{2}$ grown on HOPG cannot be associated with the structural evolution.
	
	\section{Impact of the substrate temperature on the electronic structure}
	
	Even though, the surface electronic and crystal structure of VSe$_{2}$ samples in thin films and bulk form have been extensively studied \cite{duvjir2018emergence, feng2018electronic, coelho2019charge}, the formation of the kink, the QP peak, and clear vHs points were not seen in previous studies. The discrepancy between the present work and earlier studies suggests that the growth parameters are critical in these films, close to a phase transition. Interestingly, this is analogous to the emergence of the high-T$_{c}$ superconductivity in monolayer FeSe whose Fermi surface topology is greatly affected by the specific annealing procedure \cite{he2013phase}. This process boosts the T$_{c}$ for the mono layer FeSe nearly six times compared to the bulk FeSe samples. Furthermore, the enhanced T$_{c}$ is observed only in FeSe films grown on SrTiO$_{3}$ substrate. Evidently, VSe$_{2}$ displays a very similar behavior, the superconducting phase being linked to the substrate type and the growth conditions. Furthermore, the growth conditions are known to affect the surface electronic structure and superconducting properties of cuprate superconductors by controlling the amount of electron or hole doping \cite{drozdov2018phase}.

	Following these premises, we investigated the role of the substrate temperature in the electronic structure of the 5 ML VSe$_{2}$ films grown on HOPG. The ARPES spectra for four samples grown at 550 K, 570 K, 600 K, and 650 K are shown in Fig. 2. The Fermi surfaces (top panels), band dispersions (middle panels) and EDCs (bottom panels) demonstrate that the superconducting phase forms in a relatively narrow temperature range.  At the optimal growth temperature of 570 K, QP peak exhibits the highest intensity, and the spectral kink is more pronounced (Fig. 2(b2) and 2(b3)). Both features weaken at lower and higher growth temperatures (Fig. 2(a3) and 2(c3)). The ellipsoidal electron pockets centered at the $\overline{M}$-point and the triangular electron pockets centered at the $\overline{K}$-points are also not well resolved in the samples grown at 550 K and 650 K (Fig. 2(a1) and 2(d1)). These observations show that the growth is critical in the formation of the surface electronic structure of VSe$_{2}$ films. The relatively narrow temperature range helps explaining why the superconductive phase has not been previously observed.
	
	\section{Temperature dependent evolution of the electronic structure and the energy gap}

	We have seen that the formation of the QP and the kink in the electronic structure of VSe$_{2}$ depend on the growth properties. Even though these features are commonly observed in superconductors, they do not necessarily prove the emergence of the superconducting state. In addition to these features, the Fermi gap and its temperature dependence are also critical for identifying the superconductivity. Fig. 3 summarizes the temperature dependences from these 5 ML VSe$_{2}$ films on HOPG.  Fig. 3(a) displays the evolution of the QP feature in the temperature range 10 to 180 K. The EDCs measured along the dashed line in Fig. 2(b2) track the evolution from a sharp coherent peak at low temperature to a broad incoherent spectral feature at high temperature. The QP weakens with increasing temperature but it can still be traced up to ~ 100 K (see Fig. 3(c)). Symmetrized EDCs are given in Fig. 3(c) to visually detect and extract the superconducting gap opening at the Fermi level. The symmetrization procedure has been used in the analysis of superconductors to remove the effect of the Fermi function19. The gap is observed as a spectral dip, whose width serves to estimate the gap size. From the symmetrized EDCs, it would appear that the gap persists up to nearly 150 K (Fig. 3(d)). This behavior of an apparent gap persisting at higher temperatures than the QP has been observed in high-T$_{c}$ superconductors, in which a pseudogap, in addition to the superconducting gap, plays a critical role in the electronic structure \cite{kondo2011disentangling}. Thus, the coexistence of the spectral kink, QP, and the Fermi gap strongly suggest the existence of the superconductivity in optimally grown VSe$_{2}$ samples and the temperature dependence of the gap and the QP indicates that the T$_{c}$ could be as high as 100 $\pm$ 5 K.

	The gap observed in the present work has been attributed to the emergence of superconductivity mainly in analogy with known superconductors' behavior. However, the origin of the gap could be due to other types of electronic modifications. In the case of VSe$_{2}$, the formation of a charge density wave (CDW) phase comes to mind even though VSe$_{2}$ is known to show CDW phase only at 1 ML \cite{chen2018unique}. In CDW, however, the gap should be observed only in some specific parts of the Fermi surfaces, where nesting conditions between parallel portions of the Fermi surface are satisfied \cite{gruner1994dynamics}. Instead, the gap observed here is highly isotropic. As shown in the EDCs of Fig. 3(i), the QP peak is present with the same intensity and location in all directions (see also Fig. 3(j)). Furthermore, the observed size of the gap in our films does not seem to be consistent with a CDW interpretation. In particular, in single layer VSe$_{2}$ the CDW gap was estimated from symmetrized ARPES spectra to be ~100 meV \cite{kondo2011disentangling}, much larger than the gap reported in this work, which is measured to be - 16 meV at 10 K. Nevertheless, $\mu$LEED patterns taken at 85 K and 300 K do not exhibit any differences indicating the absence of the structural transformation across the temperature range in which QP peak, spectral kink, and the Fermi gap form. Therefore, this observation excludes the structural origin of the gap opening.

	Another strong evidence for the superconducting gap is the observation of the Bogoliubov back-bending bands due to the particle-hole pair formation in the superconducting state \cite{matsui2003bcs}. This can be seen in the binding energy – in-plane momentum map taken at 10 K in Fig. 3(e) and its corresponding momentum integrated EDCs in Fig. 3(f). Band bending towards higher binding energy away from the Fermi momentum is marked with a green arc in the spectrum. At low temperatures, the QP is located at the Fermi wave vector where its binding energy is the lowest but towards the $\overline{\Gamma}$-point it shows a systematic shift to higher binding energy. On the other hand, the spectra taken at 180 K exhibits a flat feature at the Fermi level without any signature of band bending (Fig. 3(g) and 3(h)). This is an expected consequence of the disappearance particle-hole pairing in the normal state. Therefore, we can conclude that the gap at low temperature is indeed due to superconductivity. Unfortunately, ARPES cannot directly detect the T$_{c}$ of the superconductivity due to the coexistence of the superconducting and pseudogap as is the case in high-T$_{c}$ superconductors. Other types of measurements such as magnetic characterization or transport measurements are required for direct detection of the T$_{c}$. Nevertheless, the temperature dependence of the gap and the QP indicates that the T$_{c}$ could be as high as 100 $\pm$ 5 K where the coherence spectrum is significantly suppressed.

	\section{DISCUSSION and CONCLUSION}

	VSe$_{2}$ is a well-known TMDC whose band structure has been studied extensively. It is therefore surprising that new, qualitatively different spectra can be obtained from this material.  Apart from the narrow growth temperature window mentioned above, another possible explanation is the HOPG substrate used to grow the films. Most previous ARPES investigations have been conducted on VSe$_{2}$ films grown on Gr/SiC \cite{shi2015superconductivity, li2021observation, fu2008superconducting, kondo2011disentangling}. The relevance of the substrate choice is one of the main results of the present work, even though it is difficult to point to its exact role. Gr and HOPG are very similar 2D materials which are not likely to form a strong chemical bond with the VSe$_{2}$ overlayers, also in view of the large lattice mismatch. Neither substrates is then expected to induce a prominent change in the band structure of the VSe$_{2}$ films. Possibly a significant difference between the two cases results from different domain orientations and domain sizes. Compared to Gr/SiC, cleaved HOPG surfaces can be expected to expose multiple domains, oriented in different directions. These structural differences could also indirectly affect the impurities distribution between the VSe$_{2}$ layers, modifying the possible diffusion channels. Structural defects can induce superconductivity or enhance the T$_{c}$ of a superconductor as discussed for the high-T$_{c}$ superconductivity in graphite and semiconducting superlattices \cite{fogel2001novel, kopnin2013high}. Therefore, one can speculate on an indirect role of the substrate, inducing relevant modification in the VSe$_{2}$ films through minor structural differences.

	It is also interesting to note that the superconductivity in VSe$_{2}$ is not completely unprecedented: a recent work reports that bulk VSe$_{2}$ become superconducting under high pressure, albeit with a T$_{c}$ of only 4 K \cite{sahoo2020pressure}. Another interesting aspect of our results is that VSe$_{2}$ possibly being a CDW material is not expected to turn a superconductor, as the two phenomena are unlikely to coexist. In this respect, the role of the vHs points is probably critical.  For example, intercalation of impurities in TMDCs has been shown to shift the vHs points toward the Fermi level, effectively suppressing the CDW phase \cite{chikina2020turning}. This mechanism has been invoked to explain the enhancement of T$_{c}$ by more than an order of magnitude in Pd-intercalated TaSe$_{2}$ \cite{chikina2020turning, bhoi2016interplay, kim2018importance}. A similar increase in the T$_{c}$ was also realized in Cu-doped PdTe$_{2}$ whose origin is attributed to Cu intercalation which shifts the binding energy of the vHs points toward the Fermi level \cite{yan2015electronic}. Furthermore, the absence of the superconductivity in a similar compound, PtTe$_{2}$, was also directly linked to the different dispersion character of the vHs points compared to the parent superconducting materials \cite{kim2018importance}. These studies conclude that the binding energy of the vHs points is a necessary component of the superconductivity in TMDCs playing a critical role in controlling T$_{c}$.

	The role of the vHs points is also widely debated for the cuprate superconductors, whereby it has a prominent influence on the T$_{c}$ or even can induce superconductivity \cite{newns1992saddle, markiewicz1997survey}. The assumption behind the vHs scenario to explain the observed superconductivity is that the binding energy of the vHs points and the Fermi level are in close proximity. This can introduce strong divergence in the quasi two-dimensional density of states modifying the transport property of the material. This suggests the possibility of forming superconducting pairs at higher temperatures by pushing the vHs points to the Fermi level through doping or strain.

	The similarities between our findings and other superconductors are indisputable. Based on these analogies, we propose that the emergent superconductivity in VSe$_{2}$ is closely related to the formation of vHs points and their fine-tuning with respect to the Fermi level induced by impurity intercalation. Clearly, the complete understanding of our observations and their implications require further theoretical and experimental studies, possibly adopting different approaches. For example, the contribution of other bands located near or at the Fermi level may need further consideration to describe the superconducting properties of VSe$_{2}$ accurately. Finally, our work does not suggest a new material but more importantly, shows that a well-known non-superconducting material can be turned into a high-T$_{c}$ superconductor by manipulating its growth conditions. This is a significant step towards understanding high-T$_{c}$ superconductivity and its interplay with other states of matter.

	\section*{ACKNOWLEDGMENTS}
	
	This research used ESM (21-ID-1, 21-ID-2) beamline of the National Synchrotron Light Source II, a US Department of Energy (DOE) Office of Science User Facility operated for the DOE Office of Science by Brookhaven National Laboratory under Contract number DE-SC0012704. This work also used the resources of the Center for Functional Nanomaterials, Brookhaven National Laboratory, which is supported by the U.S. Department of Energy, Office of Basic Energy Sciences, under Contract number DE-SC0012704. The initial work on VSe$ _{2} $ film growth was supported by the University of Connecticut under the UCONN-REP (Grant No. 4626510).
	
	\section*{APPENDIX: METHODS}
	
	\subsection{Sample growth}
	
	Molecular beam epitaxial growth (MBE) technique was employed to grow VSe$_{2}$ thin films on HOPG and ML Gr/SiC substrates. HOPG substrates were first cleaved in air and rapidly loaded into a high vacuum chamber where they were annealed at 850 K for 30 minutes to clean the surface prior to film deposition. ML graphene were prepared by annealing SiC substrates to 1800 K for 15 minutes at the XPEEM/LEEM endstation of the 21-ID-2 beamline of NSLS-II.  Then, Gr/SiC substrates were loaded to MBE chamber and annealed at 950 K for 30 minutes to clean the surface. 5N purity Se and V with 99.8 $\%$ purity were used to grow the samples. An e-beam evaporator was used for V deposition while resistive heating of a ceramic crucible type evaporator was employed for Se. Deposition rates of V and Se were kept at 0.2 $ \AA $/min and 10 $ \AA $/min, respectively. Before starting the growth, sample and V are brought to growth temperatures and kept there for 2 hours to minimize the outgassing. We have realized that the slower growth rate can significantly enhance the quality of the ARPES data, likely due to the well-ordered samples. The sample were annealed under Se flux for 5 minutes at the growth temperatures of each sample. All the thin film samples studied here are 5 ML thick as estimated by using a quartz thickness monitor with a 15 $\%$ error bar. VSe$  _{2}$ bulk sample were obtained from 2dsemiconductors company with the part number of BLK-VSe$ _{2} $. Samples for ARPES and $\mu$LEED experiments were capped with 20 nm amorphous Se film before being removed from the MBE chamber.
	
	\subsection{Photoemission experiments}

	ARPES experiments and core-levels were recorded at 21-ID-1 ESM beamline of National Synchrotron Light Source II (NSLS-II) by using a DA30 Scienta electron spectrometer. Se 3d peak were recorded at 10 K sample temperature with 110 eV photons. The energy resolution during the photoemission experiments was better than 15 meV and a spot size of ~20 $\mu$m was used during the photoemission experiments. The Fermi level position was calibrated with respect to photoemission edge measured from Ag films deposited on the sample surface after the ARPES experiments. Prior to photoemission experiments, the VSe$ _{2} $ samples were annealed at 500 K for 30 minutes under the ultra-high vacuum condition to remove the Se capping layer. This annealing temperature was tested on various samples to insure it does not induce structural or chemical changes on the sample but still high enough to remove the capping layer.

	\subsection{$\mu$LEED}
	
	$\mu$LEED experiment was performed at X-ray photoemission electron microscopy/low-energy electron microscopy (XPEEM/LEEM) endstation of the ESM beamline (21-ID-2). The $\mu$LEED patterns were obtained after removing the Se capping layers by annealing the thin films samples at 500 K.
	
	\bibliographystyle{apsrev4-1} % Tell bibtex which bibliography style to use

\begin{thebibliography}{30}%
\makeatletter
\providecommand \@ifxundefined [1]{%
 \@ifx{#1\undefined}
}%
\providecommand \@ifnum [1]{%
 \ifnum #1\expandafter \@firstoftwo
 \else \expandafter \@secondoftwo
 \fi
}%
\providecommand \@ifx [1]{%
 \ifx #1\expandafter \@firstoftwo
 \else \expandafter \@secondoftwo
 \fi
}%
\providecommand \natexlab [1]{#1}%
\providecommand \enquote  [1]{``#1''}%
\providecommand \bibnamefont  [1]{#1}%
\providecommand \bibfnamefont [1]{#1}%
\providecommand \citenamefont [1]{#1}%
\providecommand \href@noop [0]{\@secondoftwo}%
\providecommand \href [0]{\begingroup \@sanitize@url \@href}%
\providecommand \@href[1]{\@@startlink{#1}\@@href}%
\providecommand \@@href[1]{\endgroup#1\@@endlink}%
\providecommand \@sanitize@url [0]{\catcode `\\12\catcode `\$12\catcode
  `\&12\catcode `\#12\catcode `\^12\catcode `\_12\catcode `\%12\relax}%
\providecommand \@@startlink[1]{}%
\providecommand \@@endlink[0]{}%
\providecommand \url  [0]{\begingroup\@sanitize@url \@url }%
\providecommand \@url [1]{\endgroup\@href {#1}{\urlprefix }}%
\providecommand \urlprefix  [0]{URL }%
\providecommand \Eprint [0]{\href }%
\providecommand \doibase [0]{http://dx.doi.org/}%
\providecommand \selectlanguage [0]{\@gobble}%
\providecommand \bibinfo  [0]{\@secondoftwo}%
\providecommand \bibfield  [0]{\@secondoftwo}%
\providecommand \translation [1]{[#1]}%
\providecommand \BibitemOpen [0]{}%
\providecommand \bibitemStop [0]{}%
\providecommand \bibitemNoStop [0]{.\EOS\space}%
\providecommand \EOS [0]{\spacefactor3000\relax}%
\providecommand \BibitemShut  [1]{\csname bibitem#1\endcsname}%
\let\auto@bib@innerbib\@empty
%</preamble>
\bibitem [{\citenamefont {Bednorz}\ and\ \citenamefont
  {M{\"u}ller}(1986)}]{bednorz1986possible}%
  \BibitemOpen
  \bibfield  {author} {\bibinfo {author} {\bibfnamefont {J.~G.}\ \bibnamefont
  {Bednorz}}\ and\ \bibinfo {author} {\bibfnamefont {K.~A.}\ \bibnamefont
  {M{\"u}ller}},\ }\href@noop {} {\bibfield  {journal} {\bibinfo  {journal}
  {Zeitschrift f{\"u}r Physik B Condensed Matter}\ }\textbf {\bibinfo {volume}
  {64}},\ \bibinfo {pages} {189} (\bibinfo {year} {1986})}\BibitemShut
  {NoStop}%
\bibitem [{\citenamefont {Devoret}\ \emph {et~al.}(2004)\citenamefont
  {Devoret}, \citenamefont {Wallraff},\ and\ \citenamefont
  {Martinis}}]{devoret2004superconducting}%
  \BibitemOpen
  \bibfield  {author} {\bibinfo {author} {\bibfnamefont {M.~H.}\ \bibnamefont
  {Devoret}}, \bibinfo {author} {\bibfnamefont {A.}~\bibnamefont {Wallraff}}, \
  and\ \bibinfo {author} {\bibfnamefont {J.~M.}\ \bibnamefont {Martinis}},\
  }\href@noop {} {\bibfield  {journal} {\bibinfo  {journal} {arXiv preprint
  cond-mat/0411174}\ } (\bibinfo {year} {2004})}\BibitemShut {NoStop}%
\bibitem [{\citenamefont {He}\ \emph {et~al.}(2013)\citenamefont {He},
  \citenamefont {He}, \citenamefont {Zhang}, \citenamefont {Zhao},
  \citenamefont {Liu}, \citenamefont {Liu}, \citenamefont {Mou}, \citenamefont
  {Ou}, \citenamefont {Wang}, \citenamefont {Li} \emph {et~al.}}]{he2013phase}%
  \BibitemOpen
  \bibfield  {author} {\bibinfo {author} {\bibfnamefont {S.}~\bibnamefont
  {He}}, \bibinfo {author} {\bibfnamefont {J.}~\bibnamefont {He}}, \bibinfo
  {author} {\bibfnamefont {W.}~\bibnamefont {Zhang}}, \bibinfo {author}
  {\bibfnamefont {L.}~\bibnamefont {Zhao}}, \bibinfo {author} {\bibfnamefont
  {D.}~\bibnamefont {Liu}}, \bibinfo {author} {\bibfnamefont {X.}~\bibnamefont
  {Liu}}, \bibinfo {author} {\bibfnamefont {D.}~\bibnamefont {Mou}}, \bibinfo
  {author} {\bibfnamefont {Y.-B.}\ \bibnamefont {Ou}}, \bibinfo {author}
  {\bibfnamefont {Q.-Y.}\ \bibnamefont {Wang}}, \bibinfo {author}
  {\bibfnamefont {Z.}~\bibnamefont {Li}},  \emph {et~al.},\ }\href@noop {}
  {\bibfield  {journal} {\bibinfo  {journal} {Nature materials}\ }\textbf
  {\bibinfo {volume} {12}},\ \bibinfo {pages} {605} (\bibinfo {year}
  {2013})}\BibitemShut {NoStop}%
\bibitem [{\citenamefont {Cao}\ \emph {et~al.}(2018)\citenamefont {Cao},
  \citenamefont {Fatemi}, \citenamefont {Fang}, \citenamefont {Watanabe},
  \citenamefont {Taniguchi}, \citenamefont {Kaxiras},\ and\ \citenamefont
  {Jarillo-Herrero}}]{cao2018unconventional}%
  \BibitemOpen
  \bibfield  {author} {\bibinfo {author} {\bibfnamefont {Y.}~\bibnamefont
  {Cao}}, \bibinfo {author} {\bibfnamefont {V.}~\bibnamefont {Fatemi}},
  \bibinfo {author} {\bibfnamefont {S.}~\bibnamefont {Fang}}, \bibinfo {author}
  {\bibfnamefont {K.}~\bibnamefont {Watanabe}}, \bibinfo {author}
  {\bibfnamefont {T.}~\bibnamefont {Taniguchi}}, \bibinfo {author}
  {\bibfnamefont {E.}~\bibnamefont {Kaxiras}}, \ and\ \bibinfo {author}
  {\bibfnamefont {P.}~\bibnamefont {Jarillo-Herrero}},\ }\href@noop {}
  {\bibfield  {journal} {\bibinfo  {journal} {Nature}\ }\textbf {\bibinfo
  {volume} {556}},\ \bibinfo {pages} {43} (\bibinfo {year} {2018})}\BibitemShut
  {NoStop}%
\bibitem [{\citenamefont {Manzeli}\ \emph {et~al.}(2017)\citenamefont
  {Manzeli}, \citenamefont {Ovchinnikov}, \citenamefont {Pasquier},
  \citenamefont {Yazyev},\ and\ \citenamefont {Kis}}]{manzeli20172d}%
  \BibitemOpen
  \bibfield  {author} {\bibinfo {author} {\bibfnamefont {S.}~\bibnamefont
  {Manzeli}}, \bibinfo {author} {\bibfnamefont {D.}~\bibnamefont
  {Ovchinnikov}}, \bibinfo {author} {\bibfnamefont {D.}~\bibnamefont
  {Pasquier}}, \bibinfo {author} {\bibfnamefont {O.~V.}\ \bibnamefont
  {Yazyev}}, \ and\ \bibinfo {author} {\bibfnamefont {A.}~\bibnamefont {Kis}},\
  }\href@noop {} {\bibfield  {journal} {\bibinfo  {journal} {Nature Reviews
  Materials}\ }\textbf {\bibinfo {volume} {2}},\ \bibinfo {pages} {1} (\bibinfo
  {year} {2017})}\BibitemShut {NoStop}%
\bibitem [{\citenamefont {Choi}\ \emph {et~al.}(2017)\citenamefont {Choi},
  \citenamefont {Choudhary}, \citenamefont {Han}, \citenamefont {Park},
  \citenamefont {Akinwande},\ and\ \citenamefont {Lee}}]{choi2017recent}%
  \BibitemOpen
  \bibfield  {author} {\bibinfo {author} {\bibfnamefont {W.}~\bibnamefont
  {Choi}}, \bibinfo {author} {\bibfnamefont {N.}~\bibnamefont {Choudhary}},
  \bibinfo {author} {\bibfnamefont {G.~H.}\ \bibnamefont {Han}}, \bibinfo
  {author} {\bibfnamefont {J.}~\bibnamefont {Park}}, \bibinfo {author}
  {\bibfnamefont {D.}~\bibnamefont {Akinwande}}, \ and\ \bibinfo {author}
  {\bibfnamefont {Y.~H.}\ \bibnamefont {Lee}},\ }\href@noop {} {\bibfield
  {journal} {\bibinfo  {journal} {Materials Today}\ }\textbf {\bibinfo {volume}
  {20}},\ \bibinfo {pages} {116} (\bibinfo {year} {2017})}\BibitemShut
  {NoStop}%
\bibitem [{\citenamefont {Shi}\ \emph {et~al.}(2015)\citenamefont {Shi},
  \citenamefont {Ye}, \citenamefont {Zhang}, \citenamefont {Suzuki},
  \citenamefont {Yoshida}, \citenamefont {Miyazaki}, \citenamefont {Inoue},
  \citenamefont {Saito},\ and\ \citenamefont
  {Iwasa}}]{shi2015superconductivity}%
  \BibitemOpen
  \bibfield  {author} {\bibinfo {author} {\bibfnamefont {W.}~\bibnamefont
  {Shi}}, \bibinfo {author} {\bibfnamefont {J.}~\bibnamefont {Ye}}, \bibinfo
  {author} {\bibfnamefont {Y.}~\bibnamefont {Zhang}}, \bibinfo {author}
  {\bibfnamefont {R.}~\bibnamefont {Suzuki}}, \bibinfo {author} {\bibfnamefont
  {M.}~\bibnamefont {Yoshida}}, \bibinfo {author} {\bibfnamefont
  {J.}~\bibnamefont {Miyazaki}}, \bibinfo {author} {\bibfnamefont
  {N.}~\bibnamefont {Inoue}}, \bibinfo {author} {\bibfnamefont
  {Y.}~\bibnamefont {Saito}}, \ and\ \bibinfo {author} {\bibfnamefont
  {Y.}~\bibnamefont {Iwasa}},\ }\href@noop {} {\bibfield  {journal} {\bibinfo
  {journal} {Scientific reports}\ }\textbf {\bibinfo {volume} {5}},\ \bibinfo
  {pages} {1} (\bibinfo {year} {2015})}\BibitemShut {NoStop}%
\bibitem [{\citenamefont {Li}\ \emph {et~al.}(2021)\citenamefont {Li},
  \citenamefont {Zheng}, \citenamefont {Fang}, \citenamefont {Zhang},
  \citenamefont {Chen}, \citenamefont {Chen}, \citenamefont {Liang},
  \citenamefont {Shi}, \citenamefont {Pei}, \citenamefont {Xu} \emph
  {et~al.}}]{li2021observation}%
  \BibitemOpen
  \bibfield  {author} {\bibinfo {author} {\bibfnamefont {Y.~W.}\ \bibnamefont
  {Li}}, \bibinfo {author} {\bibfnamefont {H.~J.}\ \bibnamefont {Zheng}},
  \bibinfo {author} {\bibfnamefont {Y.~Q.}\ \bibnamefont {Fang}}, \bibinfo
  {author} {\bibfnamefont {D.~Q.}\ \bibnamefont {Zhang}}, \bibinfo {author}
  {\bibfnamefont {Y.~J.}\ \bibnamefont {Chen}}, \bibinfo {author}
  {\bibfnamefont {C.}~\bibnamefont {Chen}}, \bibinfo {author} {\bibfnamefont
  {A.~J.}\ \bibnamefont {Liang}}, \bibinfo {author} {\bibfnamefont {W.~J.}\
  \bibnamefont {Shi}}, \bibinfo {author} {\bibfnamefont {D.}~\bibnamefont
  {Pei}}, \bibinfo {author} {\bibfnamefont {L.~X.}\ \bibnamefont {Xu}},  \emph
  {et~al.},\ }\href@noop {} {\bibfield  {journal} {\bibinfo  {journal} {Nature
  communications}\ }\textbf {\bibinfo {volume} {12}},\ \bibinfo {pages} {1}
  (\bibinfo {year} {2021})}\BibitemShut {NoStop}%
\bibitem [{\citenamefont {Fu}\ and\ \citenamefont
  {Kane}(2008)}]{fu2008superconducting}%
  \BibitemOpen
  \bibfield  {author} {\bibinfo {author} {\bibfnamefont {L.}~\bibnamefont
  {Fu}}\ and\ \bibinfo {author} {\bibfnamefont {C.~L.}\ \bibnamefont {Kane}},\
  }\href@noop {} {\bibfield  {journal} {\bibinfo  {journal} {Physical review
  letters}\ }\textbf {\bibinfo {volume} {100}},\ \bibinfo {pages} {096407}
  (\bibinfo {year} {2008})}\BibitemShut {NoStop}%
\bibitem [{\citenamefont {Lian}\ \emph {et~al.}(2018)\citenamefont {Lian},
  \citenamefont {Sun}, \citenamefont {Vaezi}, \citenamefont {Qi},\ and\
  \citenamefont {Zhang}}]{lian2018topological}%
  \BibitemOpen
  \bibfield  {author} {\bibinfo {author} {\bibfnamefont {B.}~\bibnamefont
  {Lian}}, \bibinfo {author} {\bibfnamefont {X.-Q.}\ \bibnamefont {Sun}},
  \bibinfo {author} {\bibfnamefont {A.}~\bibnamefont {Vaezi}}, \bibinfo
  {author} {\bibfnamefont {X.-L.}\ \bibnamefont {Qi}}, \ and\ \bibinfo {author}
  {\bibfnamefont {S.-C.}\ \bibnamefont {Zhang}},\ }\href@noop {} {\bibfield
  {journal} {\bibinfo  {journal} {Proceedings of the National Academy of
  Sciences}\ }\textbf {\bibinfo {volume} {115}},\ \bibinfo {pages} {10938}
  (\bibinfo {year} {2018})}\BibitemShut {NoStop}%
\bibitem [{\citenamefont {Fang}\ \emph {et~al.}(2019)\citenamefont {Fang},
  \citenamefont {Pan}, \citenamefont {Zhang}, \citenamefont {Wang},
  \citenamefont {Hirose}, \citenamefont {Terashima}, \citenamefont {Uji},
  \citenamefont {Yuan}, \citenamefont {Li}, \citenamefont {Tian} \emph
  {et~al.}}]{fang2019discovery}%
  \BibitemOpen
  \bibfield  {author} {\bibinfo {author} {\bibfnamefont {Y.}~\bibnamefont
  {Fang}}, \bibinfo {author} {\bibfnamefont {J.}~\bibnamefont {Pan}}, \bibinfo
  {author} {\bibfnamefont {D.}~\bibnamefont {Zhang}}, \bibinfo {author}
  {\bibfnamefont {D.}~\bibnamefont {Wang}}, \bibinfo {author} {\bibfnamefont
  {H.~T.}\ \bibnamefont {Hirose}}, \bibinfo {author} {\bibfnamefont
  {T.}~\bibnamefont {Terashima}}, \bibinfo {author} {\bibfnamefont
  {S.}~\bibnamefont {Uji}}, \bibinfo {author} {\bibfnamefont {Y.}~\bibnamefont
  {Yuan}}, \bibinfo {author} {\bibfnamefont {W.}~\bibnamefont {Li}}, \bibinfo
  {author} {\bibfnamefont {Z.}~\bibnamefont {Tian}},  \emph {et~al.},\
  }\href@noop {} {\bibfield  {journal} {\bibinfo  {journal} {Advanced
  Materials}\ }\textbf {\bibinfo {volume} {31}},\ \bibinfo {pages} {1901942}
  (\bibinfo {year} {2019})}\BibitemShut {NoStop}%
\bibitem [{\citenamefont {Matsui}\ \emph {et~al.}(2003)\citenamefont {Matsui},
  \citenamefont {Sato}, \citenamefont {Takahashi}, \citenamefont {Wang},
  \citenamefont {Yang}, \citenamefont {Ding}, \citenamefont {Fujii},
  \citenamefont {Watanabe},\ and\ \citenamefont {Matsuda}}]{matsui2003bcs}%
  \BibitemOpen
  \bibfield  {author} {\bibinfo {author} {\bibfnamefont {H.}~\bibnamefont
  {Matsui}}, \bibinfo {author} {\bibfnamefont {T.}~\bibnamefont {Sato}},
  \bibinfo {author} {\bibfnamefont {T.}~\bibnamefont {Takahashi}}, \bibinfo
  {author} {\bibfnamefont {S.~C.}\ \bibnamefont {Wang}}, \bibinfo {author}
  {\bibfnamefont {H.~B.}\ \bibnamefont {Yang}}, \bibinfo {author}
  {\bibfnamefont {H.}~\bibnamefont {Ding}}, \bibinfo {author} {\bibfnamefont
  {T.}~\bibnamefont {Fujii}}, \bibinfo {author} {\bibfnamefont
  {T.}~\bibnamefont {Watanabe}}, \ and\ \bibinfo {author} {\bibfnamefont
  {A.}~\bibnamefont {Matsuda}},\ }\href@noop {} {\bibfield  {journal} {\bibinfo
   {journal} {Physical review letters}\ }\textbf {\bibinfo {volume} {90}},\
  \bibinfo {pages} {217002} (\bibinfo {year} {2003})}\BibitemShut {NoStop}%
\bibitem [{\citenamefont {Duvjir}\ \emph {et~al.}(2018)\citenamefont {Duvjir},
  \citenamefont {Choi}, \citenamefont {Jang}, \citenamefont {Ulstrup},
  \citenamefont {Kang}, \citenamefont {Thi~Ly}, \citenamefont {Kim},
  \citenamefont {Choi}, \citenamefont {Jozwiak}, \citenamefont {Bostwick} \emph
  {et~al.}}]{duvjir2018emergence}%
  \BibitemOpen
  \bibfield  {author} {\bibinfo {author} {\bibfnamefont {G.}~\bibnamefont
  {Duvjir}}, \bibinfo {author} {\bibfnamefont {B.~K.}\ \bibnamefont {Choi}},
  \bibinfo {author} {\bibfnamefont {I.}~\bibnamefont {Jang}}, \bibinfo {author}
  {\bibfnamefont {S.}~\bibnamefont {Ulstrup}}, \bibinfo {author} {\bibfnamefont
  {S.}~\bibnamefont {Kang}}, \bibinfo {author} {\bibfnamefont {T.}~\bibnamefont
  {Thi~Ly}}, \bibinfo {author} {\bibfnamefont {S.}~\bibnamefont {Kim}},
  \bibinfo {author} {\bibfnamefont {Y.~H.}\ \bibnamefont {Choi}}, \bibinfo
  {author} {\bibfnamefont {C.}~\bibnamefont {Jozwiak}}, \bibinfo {author}
  {\bibfnamefont {A.}~\bibnamefont {Bostwick}},  \emph {et~al.},\ }\href@noop
  {} {\bibfield  {journal} {\bibinfo  {journal} {Nano letters}\ }\textbf
  {\bibinfo {volume} {18}},\ \bibinfo {pages} {5432} (\bibinfo {year}
  {2018})}\BibitemShut {NoStop}%
\bibitem [{\citenamefont {Feng}\ \emph {et~al.}(2018)\citenamefont {Feng},
  \citenamefont {Biswas}, \citenamefont {Rajan}, \citenamefont {Watson},
  \citenamefont {Mazzola}, \citenamefont {Clark}, \citenamefont {Underwood},
  \citenamefont {Markovic}, \citenamefont {McLaren}, \citenamefont {Hunter}
  \emph {et~al.}}]{feng2018electronic}%
  \BibitemOpen
  \bibfield  {author} {\bibinfo {author} {\bibfnamefont {J.}~\bibnamefont
  {Feng}}, \bibinfo {author} {\bibfnamefont {D.}~\bibnamefont {Biswas}},
  \bibinfo {author} {\bibfnamefont {A.}~\bibnamefont {Rajan}}, \bibinfo
  {author} {\bibfnamefont {M.~D.}\ \bibnamefont {Watson}}, \bibinfo {author}
  {\bibfnamefont {F.}~\bibnamefont {Mazzola}}, \bibinfo {author} {\bibfnamefont
  {O.~J.}\ \bibnamefont {Clark}}, \bibinfo {author} {\bibfnamefont
  {K.}~\bibnamefont {Underwood}}, \bibinfo {author} {\bibfnamefont
  {I.}~\bibnamefont {Markovic}}, \bibinfo {author} {\bibfnamefont
  {M.}~\bibnamefont {McLaren}}, \bibinfo {author} {\bibfnamefont
  {A.}~\bibnamefont {Hunter}},  \emph {et~al.},\ }\href@noop {} {\bibfield
  {journal} {\bibinfo  {journal} {Nano letters}\ }\textbf {\bibinfo {volume}
  {18}},\ \bibinfo {pages} {4493} (\bibinfo {year} {2018})}\BibitemShut
  {NoStop}%
\bibitem [{\citenamefont {Coelho}\ \emph {et~al.}(2019)\citenamefont {Coelho},
  \citenamefont {Nguyen~Cong}, \citenamefont {Bonilla}, \citenamefont
  {Kolekar}, \citenamefont {Phan}, \citenamefont {Avila}, \citenamefont
  {Asensio}, \citenamefont {Oleynik},\ and\ \citenamefont
  {Batzill}}]{coelho2019charge}%
  \BibitemOpen
  \bibfield  {author} {\bibinfo {author} {\bibfnamefont {P.~M.}\ \bibnamefont
  {Coelho}}, \bibinfo {author} {\bibfnamefont {K.}~\bibnamefont {Nguyen~Cong}},
  \bibinfo {author} {\bibfnamefont {M.}~\bibnamefont {Bonilla}}, \bibinfo
  {author} {\bibfnamefont {S.}~\bibnamefont {Kolekar}}, \bibinfo {author}
  {\bibfnamefont {M.-H.}\ \bibnamefont {Phan}}, \bibinfo {author}
  {\bibfnamefont {J.}~\bibnamefont {Avila}}, \bibinfo {author} {\bibfnamefont
  {M.~C.}\ \bibnamefont {Asensio}}, \bibinfo {author} {\bibfnamefont {I.~I.}\
  \bibnamefont {Oleynik}}, \ and\ \bibinfo {author} {\bibfnamefont
  {M.}~\bibnamefont {Batzill}},\ }\href@noop {} {\bibfield  {journal} {\bibinfo
   {journal} {The Journal of Physical Chemistry C}\ }\textbf {\bibinfo {volume}
  {123}},\ \bibinfo {pages} {14089} (\bibinfo {year} {2019})}\BibitemShut
  {NoStop}%
\bibitem [{\citenamefont {Dessau}\ \emph {et~al.}(1991)\citenamefont {Dessau},
  \citenamefont {Wells}, \citenamefont {Shen}, \citenamefont {Spicer},
  \citenamefont {Arko}, \citenamefont {List}, \citenamefont {Mitzi},\ and\
  \citenamefont {Kapitulnik}}]{dessau1991anomalous}%
  \BibitemOpen
  \bibfield  {author} {\bibinfo {author} {\bibfnamefont {D.~S.}\ \bibnamefont
  {Dessau}}, \bibinfo {author} {\bibfnamefont {B.~O.}\ \bibnamefont {Wells}},
  \bibinfo {author} {\bibfnamefont {Z.~X.}\ \bibnamefont {Shen}}, \bibinfo
  {author} {\bibfnamefont {W.~E.}\ \bibnamefont {Spicer}}, \bibinfo {author}
  {\bibfnamefont {A.~J.}\ \bibnamefont {Arko}}, \bibinfo {author}
  {\bibfnamefont {R.~S.}\ \bibnamefont {List}}, \bibinfo {author}
  {\bibfnamefont {D.~B.}\ \bibnamefont {Mitzi}}, \ and\ \bibinfo {author}
  {\bibfnamefont {A.}~\bibnamefont {Kapitulnik}},\ }\href@noop {} {\bibfield
  {journal} {\bibinfo  {journal} {Physical review letters}\ }\textbf {\bibinfo
  {volume} {66}},\ \bibinfo {pages} {2160} (\bibinfo {year}
  {1991})}\BibitemShut {NoStop}%
\bibitem [{\citenamefont {Bonilla}\ \emph {et~al.}(2020)\citenamefont
  {Bonilla}, \citenamefont {Kolekar}, \citenamefont {Li}, \citenamefont {Xin},
  \citenamefont {Coelho}, \citenamefont {Lasek}, \citenamefont {Zberecki},
  \citenamefont {Lizzit}, \citenamefont {Tosi}, \citenamefont {Lacovig} \emph
  {et~al.}}]{bonilla2020compositional}%
  \BibitemOpen
  \bibfield  {author} {\bibinfo {author} {\bibfnamefont {M.}~\bibnamefont
  {Bonilla}}, \bibinfo {author} {\bibfnamefont {S.}~\bibnamefont {Kolekar}},
  \bibinfo {author} {\bibfnamefont {J.}~\bibnamefont {Li}}, \bibinfo {author}
  {\bibfnamefont {Y.}~\bibnamefont {Xin}}, \bibinfo {author} {\bibfnamefont
  {P.~M.}\ \bibnamefont {Coelho}}, \bibinfo {author} {\bibfnamefont
  {K.}~\bibnamefont {Lasek}}, \bibinfo {author} {\bibfnamefont
  {K.}~\bibnamefont {Zberecki}}, \bibinfo {author} {\bibfnamefont
  {D.}~\bibnamefont {Lizzit}}, \bibinfo {author} {\bibfnamefont
  {E.}~\bibnamefont {Tosi}}, \bibinfo {author} {\bibfnamefont {P.}~\bibnamefont
  {Lacovig}},  \emph {et~al.},\ }\href@noop {} {\bibfield  {journal} {\bibinfo
  {journal} {Advanced Materials Interfaces}\ }\textbf {\bibinfo {volume} {7}},\
  \bibinfo {pages} {2000497} (\bibinfo {year} {2020})}\BibitemShut {NoStop}%
\bibitem [{\citenamefont {Chikina}\ \emph {et~al.}(2020)\citenamefont
  {Chikina}, \citenamefont {Fedorov}, \citenamefont {Bhoi}, \citenamefont
  {Voroshnin}, \citenamefont {Haubold}, \citenamefont {Kushnirenko},
  \citenamefont {Kim},\ and\ \citenamefont {Borisenko}}]{chikina2020turning}%
  \BibitemOpen
  \bibfield  {author} {\bibinfo {author} {\bibfnamefont {A.}~\bibnamefont
  {Chikina}}, \bibinfo {author} {\bibfnamefont {A.}~\bibnamefont {Fedorov}},
  \bibinfo {author} {\bibfnamefont {D.}~\bibnamefont {Bhoi}}, \bibinfo {author}
  {\bibfnamefont {V.}~\bibnamefont {Voroshnin}}, \bibinfo {author}
  {\bibfnamefont {E.}~\bibnamefont {Haubold}}, \bibinfo {author} {\bibfnamefont
  {Y.}~\bibnamefont {Kushnirenko}}, \bibinfo {author} {\bibfnamefont {K.~H.}\
  \bibnamefont {Kim}}, \ and\ \bibinfo {author} {\bibfnamefont
  {S.}~\bibnamefont {Borisenko}},\ }\href@noop {} {\bibfield  {journal}
  {\bibinfo  {journal} {npj Quantum Materials}\ }\textbf {\bibinfo {volume}
  {5}},\ \bibinfo {pages} {1} (\bibinfo {year} {2020})}\BibitemShut {NoStop}%
\bibitem [{\citenamefont {Drozdov}\ \emph {et~al.}(2018)\citenamefont
  {Drozdov}, \citenamefont {Pletikosi{\'c}}, \citenamefont {Kim}, \citenamefont
  {Fujita}, \citenamefont {Gu}, \citenamefont {Davis}, \citenamefont {Johnson},
  \citenamefont {Bo{\v{z}}ovi{\'c}},\ and\ \citenamefont
  {Valla}}]{drozdov2018phase}%
  \BibitemOpen
  \bibfield  {author} {\bibinfo {author} {\bibfnamefont {I.~K.}\ \bibnamefont
  {Drozdov}}, \bibinfo {author} {\bibfnamefont {I.}~\bibnamefont
  {Pletikosi{\'c}}}, \bibinfo {author} {\bibfnamefont {C.-K.}\ \bibnamefont
  {Kim}}, \bibinfo {author} {\bibfnamefont {K.}~\bibnamefont {Fujita}},
  \bibinfo {author} {\bibfnamefont {G.~D.}\ \bibnamefont {Gu}}, \bibinfo
  {author} {\bibfnamefont {J.~C.~S.}\ \bibnamefont {Davis}}, \bibinfo {author}
  {\bibfnamefont {P.~D.}\ \bibnamefont {Johnson}}, \bibinfo {author}
  {\bibfnamefont {I.}~\bibnamefont {Bo{\v{z}}ovi{\'c}}}, \ and\ \bibinfo
  {author} {\bibfnamefont {T.}~\bibnamefont {Valla}},\ }\href@noop {}
  {\bibfield  {journal} {\bibinfo  {journal} {Nature communications}\ }\textbf
  {\bibinfo {volume} {9}},\ \bibinfo {pages} {1} (\bibinfo {year}
  {2018})}\BibitemShut {NoStop}%
\bibitem [{\citenamefont {Kondo}\ \emph {et~al.}(2011)\citenamefont {Kondo},
  \citenamefont {Hamaya}, \citenamefont {Palczewski}, \citenamefont {Takeuchi},
  \citenamefont {Wen}, \citenamefont {Xu}, \citenamefont {Gu}, \citenamefont
  {Schmalian},\ and\ \citenamefont {Kaminski}}]{kondo2011disentangling}%
  \BibitemOpen
  \bibfield  {author} {\bibinfo {author} {\bibfnamefont {T.}~\bibnamefont
  {Kondo}}, \bibinfo {author} {\bibfnamefont {Y.}~\bibnamefont {Hamaya}},
  \bibinfo {author} {\bibfnamefont {A.~D.}\ \bibnamefont {Palczewski}},
  \bibinfo {author} {\bibfnamefont {T.}~\bibnamefont {Takeuchi}}, \bibinfo
  {author} {\bibfnamefont {J.~S.}\ \bibnamefont {Wen}}, \bibinfo {author}
  {\bibfnamefont {Z.~J.}\ \bibnamefont {Xu}}, \bibinfo {author} {\bibfnamefont
  {G.}~\bibnamefont {Gu}}, \bibinfo {author} {\bibfnamefont {J.}~\bibnamefont
  {Schmalian}}, \ and\ \bibinfo {author} {\bibfnamefont {A.}~\bibnamefont
  {Kaminski}},\ }\href@noop {} {\bibfield  {journal} {\bibinfo  {journal}
  {Nature Physics}\ }\textbf {\bibinfo {volume} {7}},\ \bibinfo {pages} {21}
  (\bibinfo {year} {2011})}\BibitemShut {NoStop}%
\bibitem [{\citenamefont {Chen}\ \emph {et~al.}(2018)\citenamefont {Chen},
  \citenamefont {Pai}, \citenamefont {Chan}, \citenamefont {Madhavan},
  \citenamefont {Chou}, \citenamefont {Mo}, \citenamefont {Fedorov},\ and\
  \citenamefont {Chiang}}]{chen2018unique}%
  \BibitemOpen
  \bibfield  {author} {\bibinfo {author} {\bibfnamefont {P.}~\bibnamefont
  {Chen}}, \bibinfo {author} {\bibfnamefont {W.~W.}\ \bibnamefont {Pai}},
  \bibinfo {author} {\bibfnamefont {Y.-H.}\ \bibnamefont {Chan}}, \bibinfo
  {author} {\bibfnamefont {V.}~\bibnamefont {Madhavan}}, \bibinfo {author}
  {\bibfnamefont {M.-Y.}\ \bibnamefont {Chou}}, \bibinfo {author}
  {\bibfnamefont {S.-K.}\ \bibnamefont {Mo}}, \bibinfo {author} {\bibfnamefont
  {A.-V.}\ \bibnamefont {Fedorov}}, \ and\ \bibinfo {author} {\bibfnamefont
  {T.-C.}\ \bibnamefont {Chiang}},\ }\href@noop {} {\bibfield  {journal}
  {\bibinfo  {journal} {Physical review letters}\ }\textbf {\bibinfo {volume}
  {121}},\ \bibinfo {pages} {196402} (\bibinfo {year} {2018})}\BibitemShut
  {NoStop}%
\bibitem [{\citenamefont {Gr{\"u}ner}(1994)}]{gruner1994dynamics}%
  \BibitemOpen
  \bibfield  {author} {\bibinfo {author} {\bibfnamefont {G.}~\bibnamefont
  {Gr{\"u}ner}},\ }\href@noop {} {\bibfield  {journal} {\bibinfo  {journal}
  {Reviews of modern physics}\ }\textbf {\bibinfo {volume} {66}},\ \bibinfo
  {pages} {1} (\bibinfo {year} {1994})}\BibitemShut {NoStop}%
\bibitem [{\citenamefont {Fogel}\ \emph {et~al.}(2001)\citenamefont {Fogel},
  \citenamefont {Pokhila}, \citenamefont {Bomze}, \citenamefont {Sipatov},
  \citenamefont {Fedorenko},\ and\ \citenamefont {Shekhter}}]{fogel2001novel}%
  \BibitemOpen
  \bibfield  {author} {\bibinfo {author} {\bibfnamefont {N.~Y.}\ \bibnamefont
  {Fogel}}, \bibinfo {author} {\bibfnamefont {A.~S.}\ \bibnamefont {Pokhila}},
  \bibinfo {author} {\bibfnamefont {Y.~V.}\ \bibnamefont {Bomze}}, \bibinfo
  {author} {\bibfnamefont {A.~Y.}\ \bibnamefont {Sipatov}}, \bibinfo {author}
  {\bibfnamefont {A.~I.}\ \bibnamefont {Fedorenko}}, \ and\ \bibinfo {author}
  {\bibfnamefont {R.~I.}\ \bibnamefont {Shekhter}},\ }\href@noop {} {\bibfield
  {journal} {\bibinfo  {journal} {Physical review letters}\ }\textbf {\bibinfo
  {volume} {86}},\ \bibinfo {pages} {512} (\bibinfo {year} {2001})}\BibitemShut
  {NoStop}%
\bibitem [{\citenamefont {Kopnin}\ \emph {et~al.}(2013)\citenamefont {Kopnin},
  \citenamefont {Ij{\"a}s}, \citenamefont {Harju},\ and\ \citenamefont
  {Heikkil{\"a}}}]{kopnin2013high}%
  \BibitemOpen
  \bibfield  {author} {\bibinfo {author} {\bibfnamefont {N.~B.}\ \bibnamefont
  {Kopnin}}, \bibinfo {author} {\bibfnamefont {M.}~\bibnamefont {Ij{\"a}s}},
  \bibinfo {author} {\bibfnamefont {A.}~\bibnamefont {Harju}}, \ and\ \bibinfo
  {author} {\bibfnamefont {T.~T.}\ \bibnamefont {Heikkil{\"a}}},\ }\href@noop
  {} {\bibfield  {journal} {\bibinfo  {journal} {Physical Review B}\ }\textbf
  {\bibinfo {volume} {87}},\ \bibinfo {pages} {140503(R)} (\bibinfo {year}
  {2013})}\BibitemShut {NoStop}%
\bibitem [{\citenamefont {Sahoo}\ \emph {et~al.}(2020)\citenamefont {Sahoo},
  \citenamefont {Dutta}, \citenamefont {Harnagea}, \citenamefont {Sood},\ and\
  \citenamefont {Karmakar}}]{sahoo2020pressure}%
  \BibitemOpen
  \bibfield  {author} {\bibinfo {author} {\bibfnamefont {S.}~\bibnamefont
  {Sahoo}}, \bibinfo {author} {\bibfnamefont {U.}~\bibnamefont {Dutta}},
  \bibinfo {author} {\bibfnamefont {L.}~\bibnamefont {Harnagea}}, \bibinfo
  {author} {\bibfnamefont {A.~K.}\ \bibnamefont {Sood}}, \ and\ \bibinfo
  {author} {\bibfnamefont {S.}~\bibnamefont {Karmakar}},\ }\href@noop {}
  {\bibfield  {journal} {\bibinfo  {journal} {Physical Review B}\ }\textbf
  {\bibinfo {volume} {101}},\ \bibinfo {pages} {014514} (\bibinfo {year}
  {2020})}\BibitemShut {NoStop}%
\bibitem [{\citenamefont {Bhoi}\ \emph {et~al.}(2016)\citenamefont {Bhoi},
  \citenamefont {Khim}, \citenamefont {Nam}, \citenamefont {Lee}, \citenamefont
  {Kim}, \citenamefont {Jeon}, \citenamefont {Min}, \citenamefont {Park},\ and\
  \citenamefont {Kim}}]{bhoi2016interplay}%
  \BibitemOpen
  \bibfield  {author} {\bibinfo {author} {\bibfnamefont {D.}~\bibnamefont
  {Bhoi}}, \bibinfo {author} {\bibfnamefont {S.}~\bibnamefont {Khim}}, \bibinfo
  {author} {\bibfnamefont {W.}~\bibnamefont {Nam}}, \bibinfo {author}
  {\bibfnamefont {B.~S.}\ \bibnamefont {Lee}}, \bibinfo {author} {\bibfnamefont
  {C.}~\bibnamefont {Kim}}, \bibinfo {author} {\bibfnamefont {B.-G.}\
  \bibnamefont {Jeon}}, \bibinfo {author} {\bibfnamefont {B.~H.}\ \bibnamefont
  {Min}}, \bibinfo {author} {\bibfnamefont {S.}~\bibnamefont {Park}}, \ and\
  \bibinfo {author} {\bibfnamefont {K.~H.}\ \bibnamefont {Kim}},\ }\href@noop
  {} {\bibfield  {journal} {\bibinfo  {journal} {Scientific reports}\ }\textbf
  {\bibinfo {volume} {6}},\ \bibinfo {pages} {1} (\bibinfo {year}
  {2016})}\BibitemShut {NoStop}%
\bibitem [{\citenamefont {Kim}\ \emph {et~al.}(2018)\citenamefont {Kim},
  \citenamefont {Kim}, \citenamefont {Kim}, \citenamefont {Kim}, \citenamefont
  {Park},\ and\ \citenamefont {Min}}]{kim2018importance}%
  \BibitemOpen
  \bibfield  {author} {\bibinfo {author} {\bibfnamefont {K.}~\bibnamefont
  {Kim}}, \bibinfo {author} {\bibfnamefont {S.}~\bibnamefont {Kim}}, \bibinfo
  {author} {\bibfnamefont {J.~S.}\ \bibnamefont {Kim}}, \bibinfo {author}
  {\bibfnamefont {H.}~\bibnamefont {Kim}}, \bibinfo {author} {\bibfnamefont
  {J.-H.}\ \bibnamefont {Park}}, \ and\ \bibinfo {author} {\bibfnamefont
  {B.~I.}\ \bibnamefont {Min}},\ }\href@noop {} {\bibfield  {journal} {\bibinfo
   {journal} {Physical Review B}\ }\textbf {\bibinfo {volume} {97}},\ \bibinfo
  {pages} {165102} (\bibinfo {year} {2018})}\BibitemShut {NoStop}%
\bibitem [{\citenamefont {Yan}\ \emph {et~al.}(2015)\citenamefont {Yan},
  \citenamefont {Jian-Zhou}, \citenamefont {Li}, \citenamefont {Cheng-Tian},
  \citenamefont {Cheng}, \citenamefont {De-Fa}, \citenamefont {Ying-Ying},
  \citenamefont {Zhuo-Jin}, \citenamefont {Jun-Feng}, \citenamefont {Chao-Yu}
  \emph {et~al.}}]{yan2015electronic}%
  \BibitemOpen
  \bibfield  {author} {\bibinfo {author} {\bibfnamefont {L.}~\bibnamefont
  {Yan}}, \bibinfo {author} {\bibfnamefont {Z.}~\bibnamefont {Jian-Zhou}},
  \bibinfo {author} {\bibfnamefont {Y.}~\bibnamefont {Li}}, \bibinfo {author}
  {\bibfnamefont {L.}~\bibnamefont {Cheng-Tian}}, \bibinfo {author}
  {\bibfnamefont {H.}~\bibnamefont {Cheng}}, \bibinfo {author} {\bibfnamefont
  {L.}~\bibnamefont {De-Fa}}, \bibinfo {author} {\bibfnamefont
  {P.}~\bibnamefont {Ying-Ying}}, \bibinfo {author} {\bibfnamefont
  {X.}~\bibnamefont {Zhuo-Jin}}, \bibinfo {author} {\bibfnamefont
  {H.}~\bibnamefont {Jun-Feng}}, \bibinfo {author} {\bibfnamefont
  {C.}~\bibnamefont {Chao-Yu}},  \emph {et~al.},\ }\href@noop {} {\bibfield
  {journal} {\bibinfo  {journal} {Chinese Physics B}\ }\textbf {\bibinfo
  {volume} {24}},\ \bibinfo {pages} {067401} (\bibinfo {year}
  {2015})}\BibitemShut {NoStop}%
\bibitem [{\citenamefont {Newns}\ \emph {et~al.}(1992)\citenamefont {Newns},
  \citenamefont {Krishnamurthy}, \citenamefont {Pattnaik}, \citenamefont
  {Tsuei},\ and\ \citenamefont {Kane}}]{newns1992saddle}%
  \BibitemOpen
  \bibfield  {author} {\bibinfo {author} {\bibfnamefont {D.~M.}\ \bibnamefont
  {Newns}}, \bibinfo {author} {\bibfnamefont {H.~R.}\ \bibnamefont
  {Krishnamurthy}}, \bibinfo {author} {\bibfnamefont {P.~C.}\ \bibnamefont
  {Pattnaik}}, \bibinfo {author} {\bibfnamefont {C.~C.}\ \bibnamefont {Tsuei}},
  \ and\ \bibinfo {author} {\bibfnamefont {C.~L.}\ \bibnamefont {Kane}},\
  }\href@noop {} {\bibfield  {journal} {\bibinfo  {journal} {Physical review
  letters}\ }\textbf {\bibinfo {volume} {69}},\ \bibinfo {pages} {1264}
  (\bibinfo {year} {1992})}\BibitemShut {NoStop}%
\bibitem [{\citenamefont {Markiewicz}(1997)}]{markiewicz1997survey}%
  \BibitemOpen
  \bibfield  {author} {\bibinfo {author} {\bibfnamefont {R.~S.}\ \bibnamefont
  {Markiewicz}},\ }\href@noop {} {\bibfield  {journal} {\bibinfo  {journal}
  {Journal of Physics and Chemistry of Solids}\ }\textbf {\bibinfo {volume}
  {58}},\ \bibinfo {pages} {1179} (\bibinfo {year} {1997})}\BibitemShut
  {NoStop}%
\end{thebibliography}%

\end{document}